\documentclass[aps,prl,twocolumn,amsmath,amssymb,showpacs,showkeys]{revtex4}
\usepackage{bm}
\newtheorem{theorem}{Theorem}
\newtheorem{remark}{Remark}

\begin{document}


\title{A teleparallel model for the neutrino}


\author{Dmitri Vassiliev}
\email[]{D.Vassiliev@bath.ac.uk}
\homepage[]{www.bath.ac.uk/~masdv/}
\affiliation{Department of Mathematical Sciences,
University of Bath, Bath BA2 7AY, UK}


\date{\today}

\begin{abstract}
The main result of the paper is a new representation for the Weyl
Lagrangian (massless Dirac Lagrangian). As the dynamical variable we use
the coframe, i.e. an orthonormal tetrad of covector fields.
We write down a simple Lagrangian -- wedge product of axial torsion with a
lightlike element of the coframe -- and show that variation of the
resulting action with respect to the coframe
produces the Weyl equation. The advantage of our approach
is that it does not require the use of spinors, Pauli matrices or
covariant differentiation. The only geometric concepts we use are
those of a metric, differential form, wedge product and exterior derivative.
Our result assigns a variational meaning to the tetrad
representation of the Weyl equation suggested by J.B.~Griffiths and R.A.~Newing.
\end{abstract}

\pacs{04.50.+h, 11.15.-q, 14.60.Lm}
\keywords{neutrino, spin, teleparallel, gravity}

\maketitle

\section{Main result}

Throughout this paper we work on a 4-manifold $M$ equipped with
prescribed Lorentzian metric $g$. In the following two subsections
we describe two different models for the neutrino.

\subsection{Traditional model}
The accepted mathematical model
for a neutrino field is the following linear partial differential
equation on $M$ know as the \emph{Weyl equation}:
\begin{equation}
\label{Weyl's equation}
i\sigma^\alpha{}_{a\dot b}\{\nabla\}_\alpha\xi^a=0.
\end{equation}
The corresponding Lagrangian is
\begin{equation}
\label{Weyl's Lagrangian}
L_\mathrm{Weyl}(\xi):=
\frac i2
(\bar\xi^{\dot b}\sigma^\alpha{}_{a\dot b}\{\nabla\}_\alpha\xi^a
-
\xi^a\sigma^\alpha{}_{a\dot b}\{\nabla\}_\alpha\bar\xi^{\dot b})
*1.
\end{equation}
Here
$\sigma^\alpha$, $\alpha=0,1,2,3$, are Pauli matrices,
$\xi$ is the unknown spinor field,
and $\{\nabla\}$ is the covariant derivative with
respect to the Levi-Civita connection:
$\{\nabla\}_\alpha\xi^a:=
\partial_\alpha\xi^a
+\frac14\sigma_\beta{}^{a\dot c}
(\partial_\alpha\sigma^\beta{}_{b\dot c}
+\{\Gamma\}^\beta{}_{\alpha\gamma}\sigma^\gamma{}_{b\dot c})\xi^b$
where $\{\Gamma\}^\beta{}_{\alpha\gamma}$ are Christoffel symbols
uniquely determined by the metric.

\subsection{Teleparallel model}

The purpose of our paper is to give an alternative representation
for the Weyl equation (\ref{Weyl's equation})
and the Weyl Lagrangian (\ref{Weyl's Lagrangian}).
To this end, we follow \cite{MR0332092} in
introducing instead of the spinor field a different unknown
-- the so-called \emph{coframe}. A coframe is a quartet of real
covector fields
$\vartheta^j$, $j=0,1,2,3$,
satisfying the constraint
\begin{equation}
\label{constraint for coframe}
g=o_{jk}\,\vartheta^j\otimes\vartheta^k
\end{equation}
where $o_{jk}=o^{jk}:=\operatorname{diag}(1,-1,-1,-1)$.
In other words, the coframe is a field of orthonormal bases with
orthonormality understood in the Lorentzian sense.
Of course, at every point of the manifold $M$ the choice of
coframe is not unique: there are 6 real degrees of freedom in
choosing the coframe and any pair of coframes is related by a
Lorentz transformation.

We define an affine connection and corresponding covariant
derivative $|\nabla|$ from the conditions
\begin{equation}
\label{defining condition for connection}
|\nabla|\vartheta^j=0.
\end{equation}
Let us emphasize that we follow \cite{MR1841284,MR1925542,MR2132536,MR2176749}
in employing holonomic coordinates, so in explicit form
conditions (\ref{defining condition for connection}) read
$\partial_\alpha\vartheta^j_\beta-|\Gamma|^\gamma{}_{\alpha\beta}\vartheta^j_\gamma=0$
giving a system of linear algebraic equations for the unknown
connection coefficients $|\Gamma|^\lambda{}_{\mu\nu}$.
The connection defined by
the system of equations~(\ref{defining condition for connection})
is called the \emph{teleparallel} or \emph{Weitzenb\"ock} connection.

Let $l$ be a nonvanishing real lightlike teleparallel covector field
($l\cdot l=0$, $|\nabla|l=0$).
Such a covector field can be written down explicitly as $l=l_j\vartheta^j$
where $l_j$ are real constants (components of the covector
$l$ in the basis $\vartheta^j$),
not all zero, satisfying
\begin{equation}
\label{constraint for constants}
o^{jk}l_jl_k=0.
\end{equation}
We define our Lagrangian as
\begin{equation}
\label{Lagrangian as function of vartheta and c}
L(\vartheta^j,l_j)=l_io_{jk}\,
\vartheta^i\wedge\vartheta^j\wedge d\vartheta^k
\end{equation}
where $d$ stands for the exterior derivative.
Note that
$\frac13\,o_{jk}\,\vartheta^j\wedge d\vartheta^k$
is the axial (totally antisymmetric) piece of torsion
of the teleparallel connection.
(The irreducible decomposition of torsion is described
in Appendix B.2 of~\cite{MR1340371}).
Let us emphasize that formula
(\ref{Lagrangian as function of vartheta and c})
does not explicitly involve connections or covariant derivatives.

The Lagrangian (\ref{Lagrangian as function of vartheta and c})
is a rank 4 covariant
antisymmetric tensor so it can be viewed as a 4-form and
integrated over the manifold
$M$ to give an invariantly defined action
$S(\vartheta^j,l_j):=\int L(\vartheta^j,l_j)$.
Independent variation with respect to
the coframe $\vartheta^j$ and parameters $l_j$
subject to the constraints (\ref{constraint for coframe})
and (\ref{constraint for constants})
produces a
pair of Euler--Lagrange equations which we write symbolically as
\begin{eqnarray}
\label{field equation 1}
\partial S(\vartheta^j,l_j)/\partial\vartheta^j&=&0,
\\
\label{field equation 2}
\partial S(\vartheta^j,l_j)/\partial l_j&=&0.
\end{eqnarray}

Observe now that the Lagrangian (\ref{Lagrangian as function of vartheta and c})
and constraints
(\ref{constraint for coframe}), (\ref{constraint for constants})
are invariant under rigid (i.e. with constant coefficients)
Lorentz transformations
\begin{equation}
\label{Lorentz transformation}
(\vartheta^j\,,\,l_j)\mapsto(\Lambda^j{}_k\vartheta^k\,,\,(\Lambda^{-1})^k{}_jl_k)
\end{equation}
where
$o_{jk}\Lambda^j{}_p\Lambda^k{}_q=o_{pq}$ and
$(\Lambda^{-1})^i{}_j\Lambda^j{}_k=\delta^i{}_k$.
This means that any variation of the parameters $l_j$ can be compensated
by a rigid variation of the coframe $\vartheta^j$. Hence,
the field equation (\ref{field equation 2})
is a consequence of the field equation (\ref{field equation 1}).
So further on we assume the parameters $l_j$ to be fixed
and study the field equation (\ref{field equation 1}) only.

\subsection{Equivalence of the two models}

Let us define the spinor field $\xi$ as the solution of the system
of equations
\begin{eqnarray}
\label{equation for spinor 1}
|\nabla|\xi&=&0,
\\
\label{equation for spinor 2}
\sigma_{\alpha a\dot b}\xi^a\bar\xi^{\dot b}&=&\pm l_\alpha
=\pm l_j\vartheta^j_\alpha
\end{eqnarray}
where
$|\nabla|_\alpha\xi^a:=
\partial_\alpha\xi^a
+\frac14\sigma_\beta{}^{a\dot c}
(\partial_\alpha\sigma^\beta{}_{b\dot c}
+|\Gamma|^\beta{}_{\alpha\gamma}\sigma^\gamma{}_{b\dot c})\xi^b$
and the sign is chosen so that the RHS lies
on the forward light cone.
The system (\ref{equation for spinor 1}), (\ref{equation for spinor 2})
determines the spinor field $\xi$ uniquely
up to a complex constant factor of modulus $1$.
This non-uniqueness is acceptable because we will be substituting
$\xi$ into the Weyl equation
(\ref{Weyl's equation}) and Weyl Lagrangian
(\ref{Weyl's Lagrangian})
which are both $\mathrm{U}(1)$-invariant.
We will call $\xi$ the spinor field \emph{associated}
with the coframe $\vartheta^j$.

The main result of our paper is the following

\begin{theorem}
\label{main result}
For any coframe $\vartheta^j$ we have
\begin{equation}
\label{relation between two Lagrangians}
L(\vartheta^j,l_j)=\pm4L_\mathrm{Weyl}(\xi)
\end{equation}
where $\xi$ is the associated spinor field.
The coframe satisfies the field equation
(\ref{field equation 1})
if and only if the associated spinor field satisfies
the Weyl equation (\ref{Weyl's equation}).
\end{theorem}

The sign in Eq.~(\ref{relation between two Lagrangians})
depends on the sign of the parameter~$l_0$,
on whether the covector $l=l_j\vartheta^j$ lies on the forward or backward light cone,
and on the orientation of the coframe
(eight different combinations).

The proof of Theorem~\ref{main result} is given below.
The crucial point is explained in the section on $B^2$-invariance,
whereas technicalities are handled in a separate section.
In the final section we discuss Theorem~\ref{main result}
within the context of know results from the theory of teleparallelism.

\section{Notation}

Our notation follows \cite{MR1841284,MR1925542,MR2132536,MR2176749}.
In particular, in line with the traditions of particle physics,
we use Greek letters to denote tensor (holonomic) indices.
Details of our spinor notation are given in
Appendix A of \cite{MR2176749}.

All our constructions are local. In particular, we restrict changes
of local coordinates on $M$ to those preserving orientation. This
restriction enables us to define the Hodge star $*$ in the usual
way.

We define the forward light cone as
the span of $\sigma_{\alpha a\dot b}\xi^a\bar\xi^{\dot b}$, $\xi\ne0$.
We also define
\[
\sigma_{\alpha\beta ac}:=(1/2)
(
\sigma_{\alpha a\dot b}\epsilon^{\dot b\dot d}\sigma_{\beta c\dot d}
-
\sigma_{\beta a\dot b}\epsilon^{\dot b\dot d}\sigma_{\alpha c\dot d}
)\,.
\]
These ``second order'' Pauli matrices are polarized, i.e.
$*\sigma=\pm i\sigma$
depending on the choice of ``basic'' Pauli matrices
$\sigma_{\alpha a\dot b}\,$.
We assume that $*\sigma=+i\sigma$.

\section{Excluding parameter-dependence}

We can always perform a restricted rigid Lorentz transformation
(\ref{Lorentz transformation})
which turns an arbitrary set of parameters $l_j$ into
$l_j=(\pm1,0,0,\pm1)$. Our model
is invariant under such transformations
so it is sufficient to prove
Theorem~\ref{main result}
for this particular
choice of parameters. Moreover, by changing, if necessary, the sign of
$L(\vartheta^j,l_j)$ we can always achieve
\begin{equation}
\label{special choice of parameters}
l_j=(1,0,0,1).
\end{equation}

Further on we assume the special choice of
parameters~(\ref{special choice of parameters}) in which case our
Lagrangian~(\ref{Lagrangian as function of vartheta and c})
takes the form
\begin{multline}
\label{Lagrangian explicit in vartheta special}
L(\vartheta^j,l_j)=(\vartheta^0+\vartheta^3)\wedge
\\
(\vartheta^0\wedge d\vartheta^0
-\vartheta^1\wedge d\vartheta^1
-\vartheta^2\wedge d\vartheta^2
-\vartheta^3\wedge d\vartheta^3).
\end{multline}

\section{$B^2$-invariance}
\label{$B^2$-invariance}

The crucial step in the proof of Theorem~\ref{main result} is the
observation that our model is invariant under a certain
class of local (i.e. with variable coefficients)
Lorentz transformations of the coframe.
In order to describe these transformations it is
convenient to switch from the real coframe
$(\vartheta^0,\vartheta^1,\vartheta^2,\vartheta^3)$ to the complex
coframe $(l,m,\bar m,n)$ where
\[
l:=\vartheta^0+\vartheta^3,\quad
m:=\vartheta^1+i\vartheta^2,\quad
n:=\vartheta^0-\vartheta^3
\]
(here the definition of $l$
is in agreement with Eq.~(\ref{special choice of parameters})).
In this new notation
the Lagrangian~(\ref{Lagrangian explicit in vartheta special})
and constraint~(\ref{constraint for coframe}) take the form
\begin{equation}
\label{Lagrangian in complex form}
L(\vartheta^j,l_j)=(1/2)\ l\wedge(n\wedge dl-\bar m\wedge dm-m\wedge d\bar m),
\end{equation}
\begin{equation}
\label{constraint for coframe in complex form}
g=(1/2)(l\otimes n+n\otimes l-m\otimes\bar m-\bar m\otimes m).
\end{equation}
Let us perform the linear transformation
of the coframe
\begin{equation}
\label{gauge transformation}
\left(\begin{matrix}
l\\ m\\\bar m\\ n
\end{matrix}\right)
\mapsto
\left(\begin{matrix}
l\\ m+fl\\\bar m+\bar fl\\ n+f\bar m+\bar fm+|f|^2l
\end{matrix}\right)
\end{equation}
where $f:M\to\mathbb{C}$ is an arbitrary scalar function.
It is easy to see that both the Lagrangian~(\ref{Lagrangian in complex form})
and the constraint~(\ref{constraint for coframe in complex form})
are invariant under the
transformation~(\ref{gauge transformation}),
hence the field equation~(\ref{field equation 1})
is also invariant.

Invariance of the field equation~(\ref{field equation 1})
means that solutions come in equivalence classes:
two coframes are said to be equivalent if there exists a scalar function
$f:M\to\mathbb{C}$ such that the transformation~(\ref{gauge transformation})
maps one
coframe into the other. In order to understand the group-theoretic nature
of these equivalence classes we note that
at every point of the manifold $M$
transformations~(\ref{gauge transformation})
form a subgroup of the restricted Lorentz group.
Moreover, this is a very special subgroup:
it is the unique nontrivial abelian subgroup of the
restricted Lorentz group. Here ``unique'' is understood as ``unique up to conjugation'',
whereas  the meaning of
``nontrivial abelian subgroup'' is explained in
Appendix C of \cite{MR2132536}.
In $\mathrm{SL}(2,\mathbb{C})$ notation the subgroup
(\ref{gauge transformation}) is written, up to conjugation, as
$B^2:=\left\{
\left.
\left(
\begin{matrix}
\ 1\ &\ f\ \\
\ 0\ &\ 1\
\end{matrix}
\right)
\right|
\quad f\in\mathbb{C}
\right\}$
where the notation $B^2$ is taken from subsection 10.122 of \cite{MR867684}.
It is known that $B^2$ is the subgroup
of the restricted Lorentz group preserving a given nonzero spinor. Our equivalence
classes of coframes can be identified with cosets of $B^2$,
hence they are equivalent to spinors.

\begin{remark}
The rigorous statement is that a coset of the subgroup $B^2$
is equivalent to a spinor up to choice of sign,
i.e. spinors $\zeta$ and $-\zeta$ correspond to the same coset.
This non-uniqueness
is acceptable because it is known (see, for example, section 19 in \cite{LL4}
or section 3-5 in \cite{MR1884336})
that the sign of a spinor does not have a physical meaning.
\end{remark}

\begin{remark}
Our construction does not allow us to deal with the zero spinor.
\end{remark}

\section{Technicalities}

Arguments presented in the previous section show that
even though our field equation~(\ref{field equation 1})
has no spinors appearing in it explicitly, it is, in fact,
a first order differential equation for an unknown spinor field.
From this point it is practically inevitable
that equation~(\ref{field equation 1}) is, up to a change of variable,
Weyl's equation~(\ref{Weyl's equation}).

The actual proof of Theorem~\ref{main result} is carried out by
means of a straightforward (but lengthy) calculation. The
calculation goes as follows.

The set of coframes
has four connected components
corresponding to two different orientations, $*(l\wedge m)=\pm i(l\wedge m)$,
and to $l$ lying on the forward or backward light cone.
We assume for definiteness that we are working with
coframes satisfying $*(l\wedge m)=+i(l\wedge m)$
and with $l$ lying on the forward light cone.

It is easy to see that our transformation~(\ref{gauge transformation})
preserves the tensor $l\wedge m$. Moreover, each equivalence class of coframes
is completely determined by this tensor.
Therefore, it is convenient to identify each equivalence class
with a spinor field $\zeta$ in accordance with the formula
\begin{equation}
\label{nonlinear change of variable, new notation}
\left(l\wedge m\right)_{\alpha\beta}
=\sigma_{\alpha\beta ab}\zeta^a\zeta^b.
\end{equation}
The fact that a decomposable polarized antisymmetric tensor
is equivalent to the square of a spinor
is a standard one and
was extensively used in
\cite{MR1841284,MR1925542,MR2132536,MR2176749}.

Resolving Eq.~(\ref{nonlinear change of variable, new notation})
for the coframe $\{l,m,\bar m,n\}$, we get
the following formulas:
$l$ is given by
\begin{equation}
\label{formula for l}
l_\alpha=\sigma_{\alpha a\dot b}\zeta^a\bar\zeta^{\dot b},
\end{equation}
$n$ is an arbitrary real (co)vector field satisfying
\begin{equation}
\label{formula for n}
n\cdot n=0,\qquad l\cdot n=2,
\end{equation}
and $m$ is given by
\begin{equation}
\label{formula for m}
m_\beta=(1/2)\sigma_{\alpha\beta ab}n^\alpha\zeta^a\zeta^b.
\end{equation}

Formula~(\ref{Lagrangian in complex form}) implies
\begin{multline}
\label{formula for *L original}
*L(\vartheta^j,l_j)=(1/2)\sqrt{|\det g|}\,\varepsilon_{\alpha\beta\gamma\delta}
\\
l^\alpha(
n^\beta\{\nabla\}^\gamma l^\delta
-\bar m^\beta\{\nabla\}^\gamma m^\delta
-m^\beta\{\nabla\}^\gamma\bar m^\delta)
\end{multline}
where $\varepsilon$ is the Levi-Civita symbol,
$\varepsilon_{0123}:=+1$.
Here $\{\nabla\}$ stands for the Levi-Civita covariant derivative
which should not be confused with the teleparallel covariant
derivative $|\nabla|$.
Substituting formulas~(\ref{formula for l}) and~(\ref{formula for m})
into formula~(\ref{formula for *L original})
and using algebraic properties of
Pauli matrices as well as conditions (\ref{formula for n}),
we arrive at
\begin{equation}
\label{formula for *L}
*L(\vartheta^j,l_j)=-2i
(\bar\zeta^{\dot b}\sigma^\alpha{}_{a\dot b}\{\nabla\}_\alpha\zeta^a
-
\zeta^a\sigma^\alpha{}_{a\dot b}\{\nabla\}_\alpha\bar\zeta^{\dot b}).
\end{equation}
Formulas~(\ref{formula for *L}),~(\ref{nonlinear change of variable, new notation})
show that our Lagrangian~(\ref{Lagrangian in complex form})
is a function of $l\wedge m$
rather than of $l$ and $m$ separately. This is, of course,
a consequence of the $B^2$-invariance described in the previous section.

Applying the Hodge star to Eq.~(\ref{formula for *L})
and comparing with Eq.~(\ref{Weyl's Lagrangian}), we get
\begin{equation}
\label{relation between two Lagrangians, special}
L(\vartheta^j,l_j)=4L_\mathrm{Weyl}(\zeta).
\end{equation}

We have $|\nabla|(l\wedge m)=0$,
so formula~(\ref{nonlinear change of variable, new notation}) implies
\begin{equation}
\label{zeta is parallel}
|\nabla|\zeta=0.
\end{equation}
Comparing Eqs.~(\ref{equation for spinor 1}),~(\ref{equation for spinor 2})
with Eqs.~(\ref{zeta is parallel}),~(\ref{formula for l})
we conclude that the spinor fields $\xi$ and $\zeta$ coincide
up to a complex constant factor of modulus $1$.
The Weyl Lagrangian is $\mathrm{U}(1)$-invariant, so
in Eq.~(\ref{relation between two Lagrangians, special})
we can replace $\zeta$ by $\xi$, arriving at Eq.~(\ref{relation between two Lagrangians}).

As we have established
the identity~(\ref{relation between two Lagrangians, special})
and as
each equivalence class of coframes is equivalent to a spinor field $\zeta$,
our field equation~(\ref{field equation 1}) is equivalent to
\begin{equation}
\label{Weyl's equation, special}
i\sigma^\alpha{}_{a\dot b}\{\nabla\}_\alpha\zeta^a=0.
\end{equation}
The Weyl equation is $\mathrm{U}(1)$-invariant, so
in Eq.~(\ref{Weyl's equation, special})
we can replace $\zeta$ by $\xi$, arriving at Eq.~(\ref{Weyl's equation}).
This completes the proof of Theorem~\ref{main result}.

The detailed calculation leading to Eq.~(\ref{formula for *L})
will be presented in a separate paper.

\section{Discussion}

The subject of teleparallelism has a long history dating back to the 1920s.
Its origins lie in the pioneering works of \'E.~Cartan, A.~Einstein and R.~Weitzenb\"ock.
Modern reviews of the physics of teleparallelism are given in
\cite{MR583723,sauer-2004-,MR1661422,MR1617858}.
Note that Einstein's original papers
on the subject are now available in English translation
\cite{unzicker-2005-}.

However, the construction presented in our paper differs from the
traditional one. The crucial difference is our choice of
Lagrangian~(\ref{Lagrangian as function of vartheta and c}) which is
parameter-dependent and linear in torsion.
The vast majority of publications on the subject deal with parameter-independent
Lagrangians quadratic in torsion.
One particular parameter-independent
quadratic Lagrangian has received special attention
as it leads to a teleparallel theory of gravity
equivalent
(in terms of the resulting metric) to general relativity;
the explicit formula for this Lagrangian can be found, for example,
in \cite{MR1340371,MR1661422,MR1617858,mielke,obukhov}.

Another difference is that
in teleparallelism it is traditional to vary the coframe without any
constraints. This is because teleparallelism is usually viewed as a
framework for alternative theories of gravity and in
this setting the metric~(\ref{constraint for coframe})
has to be treated as an unknown.
We, on the other hand, vary the coframe subject to the
metric constraint~(\ref{constraint for coframe}).
This is because we view teleparallelism as a framework for
the reinterpretation of
quantum electrodynamics and in this setting the metric
plays the role of a given background.

The question of whether spin can be incorporated into the
teleparallel theory of gravity has long been the subject of debate
among specialists. The latest contributions can be found
in \cite{mielke,obukhov} with further references therein.
As we do not vary the metric our result
is not directly related to this debate but it might
motivate a fresh re-examination of the question.

It is interesting that our model exhibits similarities with
Caroll--Field--Jackiw electrodynamics
\cite{carroll,itin}.
Both involve a covariantly constant covector field:
in our model it is the lightlike covector field $l$
which is covariantly constant with respect to the teleparallel connection,
whereas in Caroll--Field--Jackiw electrodynamics it is a
timelike covector field which is
covariantly constant with respect to the Levi-Civita connection.

Our model also exhibits strong similarities with the ``bumblebee model''
discussed by V.A.~Kosteleck{\'y} \cite{kostelecky}:
our teleparallel lightlike covector field $l$ plays a role similar to that of
the ``bumblebee field''. Of course, in our case this
covector field has a simple physical interpretation:
according to Eq.~(\ref{equation for spinor 2}) it is the neutrino current.

The fact that the Weyl \emph{equation} can be rewritten in tetrad form is not
in itself new: this was done by J.B.~Griffiths and R.A.~Newing~\cite{MR0332092}.
Our new result is the tetrad representation
(\ref{Lagrangian as function of vartheta and c}) for the Weyl
\emph{Lagrangian}.

\begin{acknowledgments}
The author is grateful to F.W.~Hehl, Y.~Itin  and J.B.~Griffiths for stimulating discussions.
\end{acknowledgments}

\bibliography{weyl8}

\end{document}